\pdfoutput=1
\documentclass[conference]{IEEEtran}
\IEEEoverridecommandlockouts

\usepackage{cite}
\usepackage{amsmath,amssymb,amsfonts}
\usepackage{graphicx}
\usepackage{textcomp}
\usepackage{xcolor}
\usepackage{booktabs}
\usepackage[hyphens]{url}
\usepackage{hyperref}

\newcommand{\BraiinsPreTrucBlocks}{1{,}528}
\newcommand{\BraiinsPreTrucExp}{169}

\newcommand{\DTwoGridCells}{54}
\newcommand{\DTwoGridNotSupported}{35}

\newcommand{\DTwoShareOnePct}{16.0}
\newcommand{\DTwoVerdict}{not supported}

\newcommand{\EphDustCellsNoTFifty}{11}

\newcommand{\EphDustDTFiftyAbs}{414}

\newcommand{\FeeCodeMaxPct}{0.016}

\newcommand{\FeeOpRetManyPct}{0.024}

\newcommand{\FeeOpRetOnePct}{0.075}

\newcommand{\FeePreVthirtyDirectPct}{0.242}

\newcommand{\FeeSubOnePct}{3.1}

\newcommand{\FeeSubOneRevPct}{0.03}
\newcommand{\FeeSubOneSatBlock}{82{,}889}
\newcommand{\FeeSubOneTimesNext}{42}
\newcommand{\FeeSubsidyBtc}{3.125}

\newcommand{\FeeWindowDays}{90}

\newcommand{\FifteenDTFiftyAbs}{103}

\newcommand{\FifteenDefaultDate}{10~July~2015}

\newcommand{\FifteenEndHeight}{420{,}000}
\newcommand{\FifteenFTwoPeerRatio}{5}

\newcommand{\FifteenOptionFTwoTxs}{31}
\newcommand{\FifteenOptionTxs}{46}

\newcommand{\FifteenSettingDate}{13~February~2015}
\newcommand{\FifteenStartHeight}{300{,}000}

\newcommand{\FifteenTFiftyDate}{21~October~2015}

\newcommand{\FriendsBig}{8}

\newcommand{\FriendsNeverCode}{7}
\newcommand{\FriendsNeverCodeNoLux}{6}

\newcommand{\FriendsNeverCodeTrucExpNoLux}{1{,}835}

\newcommand{\FriendsSubOneMaxLag}{11}

\newcommand{\FrozenFloor}{3}
\newcommand{\FrozenMinTxsSubOne}{100}

\newcommand{\FrozenWindowRare}{50}
\newcommand{\FrozenWindowSubOne}{20}
\newcommand{\GridCells}{54}
\newcommand{\GridHalfCells}{27}
\newcommand{\GridMHigh}{10, 100, 500}
\newcommand{\GridMRare}{1, 2, 3}
\newcommand{\GridSList}{0.3, 0.5, 0.7}
\newcommand{\GridSMinPct}{30}
\newcommand{\GridWList}{10, 20, 50}
\newcommand{\HashWinBlocks}{4{,}320}
\newcommand{\MinBlocksPool}{100}
\newcommand{\NBlocks}{124{,}395}
\newcommand{\NPoolsCurves}{22}

\newcommand{\NodeBitTwentySixMax}{1{,}070}

\newcommand{\NodeBitTwentySixMaxKnotsPct}{99}
\newcommand{\NodeDataEndDate}{15~November~2026}

\newcommand{\NodeDaysPastBlocks}{12}
\newcommand{\NodeEndWithUa}{8{,}451}

\newcommand{\NodeEphDustDHashAbs}{86}

\newcommand{\NodeFirstDate}{15~July~2015}

\newcommand{\NodeFullRbfDCore}{273}

\newcommand{\NodeFullRbfTFiftyDate}{1~July~2025}

\newcommand{\NodeKnotsAtOpRetHashPct}{14.2}

\newcommand{\NodeKnotsMaxPct}{17.4}

\newcommand{\NodeMinersFirstShareMaxPct}{46}
\newcommand{\NodeMinersFirstShareMinPct}{0}
\newcommand{\NodeNotRelayOpRetAtHashPct}{69}
\newcommand{\NodeNotRelayOpRetEndPct}{48}

\newcommand{\NodeOnePcDHashAbs}{152}

\newcommand{\NodeOpRetOneDHashAbs}{180}

\newcommand{\NodeOpRetOneVariantsTFifty}{8}
\newcommand{\NodeOtherMaxPct}{1.2}

\newcommand{\NodePtaDHashAbs}{324}

\newcommand{\NodeRulesMinersFirst}{5}
\newcommand{\NodeRulesWave}{6}
\newcommand{\NodeSubOneAtHashPct}{0}

\newcommand{\NodeSubOneDHashAbs}{411}

\newcommand{\NodeSubOneTFiftyDate}{6~September~2026}
\newcommand{\NodeSubOneVariantsTFifty}{14}
\newcommand{\NodeTaprootAtActivationPct}{53}

\newcommand{\NodeTrucDHashAbs}{27}

\newcommand{\NodeUnmappedMaxPct}{3.5}
\newcommand{\NodeVariants}{16}
\newcommand{\NodeXcheckDates}{4}
\newcommand{\NodeXcheckExKitHighPp}{2.0}

\newcommand{\NodeXcheckKitHighPp}{5.7}

\newcommand{\NonUnitMaxSharePct}{0.69}

\newcommand{\OnePcAntDate}{13~June~2025}

\newcommand{\OnePcTFiftyDate}{14~June~2025}

\newcommand{\OpRetManyAdoptedEndPct}{67}

\newcommand{\OpRetManyDTFiftyAbs}{213}

\newcommand{\OpRetOneAdoptedEndPct}{68}

\newcommand{\OpRetOneBraiinsDate}{1~July~2026}

\newcommand{\OpRetOneDTFiftyAbs}{211}

\newcommand{\OpRetOneDTFiftyMonths}{7}

\newcommand{\OppWinBlocks}{1{,}008}
\newcommand{\OrepAfterTfiftyMaxMonth}{October~2026}
\newcommand{\OrepAfterTfiftyMaxTxs}{26{,}122}
\newcommand{\OrepAfterTfiftyMedianTxs}{8{,}466}
\newcommand{\OrepAfterVthirtyFirstMonth}{November~2025}
\newcommand{\OrepAfterVthirtyLastMonth}{January~2026}

\newcommand{\OrepAfterVthirtyMedianTxs}{68}

\newcommand{\PoolResolvedPctPrecise}{99.94}
\newcommand{\PoolUnknownBlocks}{75}

\newcommand{\PoolVerifyBlocks}{294{,}396}
\newcommand{\PoolVerifyDiffering}{0}

\newcommand{\PoolsRepoCommit}{59dc205}
\newcommand{\PreVThirtyFirstDate}{21~January~2024}

\newcommand{\PreVThirtyManyTxsDirect}{8{,}485}

\newcommand{\PreVThirtyOneTxsDirect}{12{,}016}

\newcommand{\PtaAdoptedEndPct}{89}

\newcommand{\PtaDTFiftyAbs}{606}

\newcommand{\PtaDTFiftyMonths}{20}

\newcommand{\PtaFoundryDate}{30~May~2026}

\newcommand{\PtaLtwoSpends}{4{,}871}
\newcommand{\PtaLtwoTrucZero}{4{,}587}

\newcommand{\PtaMaraDate}{12~February~2025}

\newcommand{\PtaRunesOutputPct}{99.93}
\newcommand{\PtaRunesSpends}{6{,}815{,}476}

\newcommand{\PtaTFiftyDate}{30~May~2026}

\newcommand{\RefPadBlocks}{504}

\newcommand{\RunesAntSharePct}{41.9}

\newcommand{\RunesBraiinsSharePct}{79.6}

\newcommand{\RunesBurstDate}{2~July~2026}

\newcommand{\RunesFriendsNoLuxBlocks}{1{,}298}

\newcommand{\RunesNonFriendsSharePct}{79.1}

\newcommand{\SubOneAdoptedEndPct}{100}

\newcommand{\SubOneCoreDate}{3~September~2025}
\newcommand{\SubOneCoreRelease}{v29.1}

\newcommand{\SubOneDTFiftyAbs}{43}

\newcommand{\SubOneDTFiftyMajorAbs}{79}

\newcommand{\SubOneNeverLargestBlocks}{1{,}689}
\newcommand{\SubOneNeverLargestPool}{SBI Crypto}
\newcommand{\SubOneNeverLastDate}{29~July~2026}
\newcommand{\SubOneNeverPools}{3}

\newcommand{\SubOnePR}{\#33106}

\newcommand{\SubOneTFiftyDate}{22~July~2025}

\newcommand{\TaprootActivationDate}{14~November~2021}
\newcommand{\TaprootActivationHeight}{709{,}632}
\newcommand{\TaprootBlocks}{7{,}000}

\newcommand{\TaprootDayZeroPools}{6}

\newcommand{\TaprootPools}{9}

\newcommand{\TaprootWindowEnd}{715{,}999}
\newcommand{\TaprootWindowStart}{709{,}000}
\newcommand{\TrucAdoptedEndPct}{89}

\newcommand{\TrucAntDate}{4~June~2025}

\newcommand{\TrucBraiinsDate}{30~June~2026}

\newcommand{\TrucBraiinsLagAbs}{637}

\newcommand{\TrucCoreRelease}{v28.0}

\newcommand{\TrucDTFiftyAbs}{246}

\newcommand{\TrucDTFiftyMonths}{8}

\newcommand{\TrucMaraDate}{24~April~2025}

\newcommand{\TrucNetShareMaxPreJunePct}{0.17}

\newcommand{\TrucTFiftyDate}{4~June~2025}

\newcommand{\VThirtyDate}{9~October~2025}
\newcommand{\ValBnocEdExact}{7}
\newcommand{\ValBnocEdListed}{7}
\newcommand{\ValBnocPtaExact}{8}
\newcommand{\ValBnocPtaListed}{8}
\newcommand{\ValFirstPtrHeight}{709{,}635}

\newcommand{\ValFtwoHalfOutlierTxs}{1}
\newcommand{\ValFtwoHalfTxs}{18{,}600}
\newcommand{\ValFtwoPointOneDate}{24~August~2025}

\newcommand{\ValFtwoQuarterDate}{18~August~2025}
\newcommand{\ValFtwoQuarterMinRate}{0.24}

\newcommand{\ValMononautRatioPct}{98.9}

\newcommand{\WideStartDate}{24~July~2023}
\newcommand{\WideStartHeight}{800{,}000}
\newcommand{\WindowEndDate}{3~November~2026}
\newcommand{\WindowEndHeight}{974{,}394}
\newcommand{\WindowStartDate}{29~June~2024}
\newcommand{\WindowStartHeight}{850{,}000}

\newcommand{\PaperTitle}{Who Sets Bitcoin's Mempool Policy?}
\newcommand{\PaperSubtitle}{Evidence from Blocks and Node Snapshots, 2024--2026}

\hypersetup{pdftitle={\PaperTitle{} \PaperSubtitle}, pdfauthor={Saurabh Kumar}}

\begin{document}

\title{\PaperTitle\\
{\LARGE \PaperSubtitle}}

\author{
\IEEEauthorblockN{Saurabh Kumar}
\IEEEauthorblockA{Bitcoin Data Labs\\
sorukumar@bitcoindatalabs.org}
}

\maketitle
\thispagestyle{plain}
\pagestyle{plain}

\begin{abstract}
Every Bitcoin node and mining pool decides which unconfirmed transactions to accept with a local mempool policy, and Bitcoin Core changes the defaults in its releases. Core writes the defaults, node operators choose what to run and pools choose what to mine, but no study has dated all three across a full wave of changes. We date seven changes of 2024--2026 for \NPoolsCurves{} pools across \NBlocks{} blocks, with an estimator fixed before we ran it, and set them beside daily snapshots of reachable nodes. The estimator reproduces 0xB10C's first sightings exactly and dates Taproot adoption to its activation day. The answer depends on what a change offered pools. Where it paid and needed only a setting, miners set the policy. Half the hashrate mined sub-1~sat/vB transactions \SubOneDTFiftyAbs{} days before Core's release, when no node relayed them by default. They earned adopters a share of fees \FeeSubOneTimesNext{} times that of any other rule. The four rules that protect Lightning channel exits paid almost nothing and reached half the hashrate \TrucDTFiftyMonths{} to \PtaDTFiftyMonths{} months after release. Of the \FriendsBig{} pools grouped with AntPool by template similarity, \FriendsNeverCode{} never adopted them. Large \texttt{OP\_RETURN} outputs reached half the hashrate \OpRetOneDTFiftyMonths{} months after Core allowed them, while \NodeNotRelayOpRetAtHashPct{}\% of nodes still did not relay them. For Lightning and wallet developers, a Core release is not deployment: it waits on the upgrade schedules of a few template builders.

\end{abstract}


\section{Introduction}\label{sec:intro}

Bitcoin is a payment network whose nodes run open-source software. Before a transaction is mined, every node decides whether to accept it and pass it on, under a local \emph{mempool policy} (also called relay policy, or standardness). Mining pools then build blocks from the transactions that reach them. These rules are not consensus: a block that breaks them is still valid. They decide, however, which transactions a user can get mined without a private arrangement with a pool. Bitcoin Core, the software most nodes run, changes the default rules in its releases.

The changes of 2024--2026 had real stakes. Four of them---version-3 (TRUC) transactions, one-parent-one-child (1p1c) package relay, pay-to-anchor (P2A) outputs and ephemeral dust---let Lightning channels close with zero-fee pre-signed transactions whose fee is added when the channel closes \cite{kumar2026forceclosecosts}. Such an exit can only be mined by a pool that accepts these rules. Another change cut the default fee floor tenfold, to 0.1~sat/vB \cite{core2025pr33106}. A third, the relaxed \texttt{OP\_RETURN} limits of \texttt{v30.0}, set off a public dispute over data in transactions \cite{poinsot2025opreturn,core2025v30notes}. During it, Bitcoin Knots, a derivative of Core that keeps the old limits \cite{knots2026source}, grew to \NodeKnotsMaxPct{}\% of reachable nodes (Section~\ref{sec:res-nodes}).

Who actually sets this policy? Core's maintainers write the defaults. Node operators choose which release, or which client, to run, and so which transactions they pass on. Mining pools choose what goes into blocks, through the block templates they build. A rule is in force for a user only once a pool will mine it. For each change of the 2024--2026 wave, we ask which of the three moved first, and why.

Blocks show what pools mined, not what they refused, so scarce demand hides adoption. Before June 2025, TRUC transactions appeared in at most \TrucNetShareMaxPreJunePct{}\% of blocks in any month; a pool that would mine them and one that would not looked alike. We therefore compare each pool with the rest of the network over the same blocks, with one estimator whose parameters we fixed before running it on any rule. We validate it against every published sighting we found, and against Taproot, which every upgraded pool could mine from the same block. We add daily snapshots of reachable nodes from KIT's monitor \cite{kit2026dsnmonitor} and the fees each rule's transactions paid.

This paper continues a series on how Bitcoin's open-source software reaches the people who run it: client diversity \cite{kumar2026clientdiversity} and patch uptake \cite{kumar2026upgradedynamics} among Lightning nodes, and the cost of Lightning force closes \cite{kumar2026forceclosecosts}, which depends on the rules studied here. We now turn to the base layer's transaction policy.

\begin{figure*}[t]
\centering
\includegraphics[width=\textwidth]{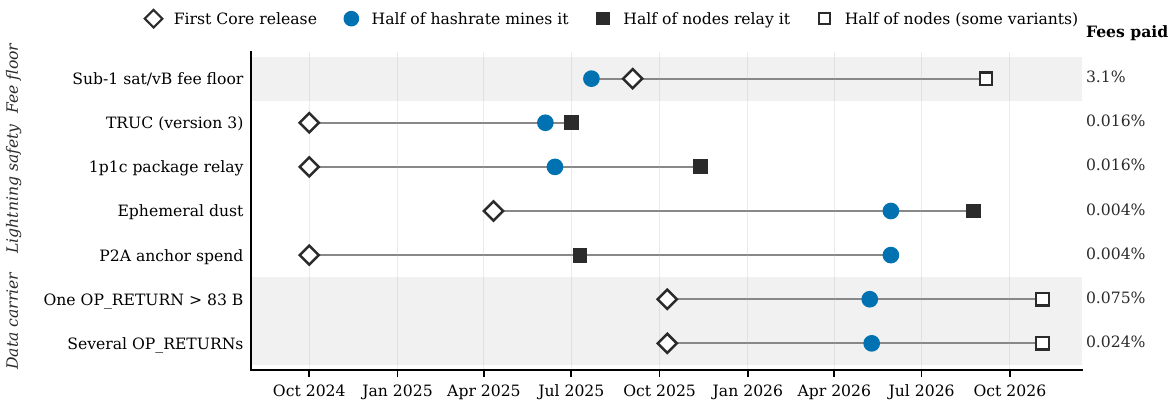}
\caption{For each rule of the 2024--2026 wave: Core's first release carrying it, the day half the hashrate mined it, and the day half of reachable nodes relayed it by default. On the right, the share of their fee income that the rule's transactions paid adopting pools in the \FeeWindowDays{} days after half the hashrate adopted. Open squares: the node date holds in only some of the \NodeVariants{} node variants. Sources: Tables~\ref{tab:thesis} and~\ref{tab:fees}.}
\label{fig:chain}
\end{figure*}

Who moved first turned on two things: whether a setting was enough, and whether the change paid (Fig.~\ref{fig:chain}, Table~\ref{tab:thesis}):
\begin{itemize}
\item \textbf{Fee floor: miners first.} Half the hashrate mined sub-1~sat/vB transactions \SubOneDTFiftyAbs{} days before Core first shipped the lower floor (\SubOneCoreRelease{}, \SubOneCoreDate{}), when no release relayed them by default. The change needed only a setting, and these transactions brought adopters \FeeSubOnePct{}\% of their fee income.
\item \textbf{Lightning safety rules: Core first, miners months later.} These rules needed new Core code and paid adopters at most \FeeCodeMaxPct{}\% of fee income. Half the hashrate mined them \TrucDTFiftyMonths{} to \PtaDTFiftyMonths{} months after release. \FriendsNeverCode{} of the \FriendsBig{} pools grouped with AntPool by template similarity never adopted any of them.
\item \textbf{Data carrier: contested.} Half the hashrate mined large \texttt{OP\_RETURN} outputs \OpRetOneDTFiftyAbs{} days after \texttt{v30.0}, and \OpRetOneAdoptedEndPct{}\% by the end of the window. A test we fixed in advance, of whether a startup setting lets pools move first, fails here.
\item \textbf{Nodes: last on most rules.} On \NodeRulesMinersFirst{} of the \NodeRulesWave{} rules with their own node defaults (both \texttt{OP\_RETURN} forms share one), half the hashrate mined the rule before half of reachable nodes relayed it by default.
\end{itemize}

\paragraph*{Contributions} We build on the monitoring of 0xB10C \cite{b10c2025bnocversions,b10c2024templates,b10c2026friends}, mononaut \cite{scoresby2025mononaut}, Towns \cite{towns2026opreturn} and others, and reproduce their published numbers. To our knowledge, none of the following has been measured before for a whole wave of changes:
\begin{enumerate}
\item Per-pool adoption dates for the full 2024--2026 wave, against three Core dates and across a \GridCells{}-cell sensitivity grid, released as a dataset.
\item The node side of every rule, including the full-RBF default that blocks cannot show.
\item The fees each rule paid the pools that adopted it, which turns ``this change pays'' from a label into a measurement.
\item A test, written down before we ran it, of whether pools lead on changes that need only a setting. It fails; we report it as a negative result, with one observation made after seeing the data.
\item Behavioral evidence for template sharing: the pools whose block templates resemble AntPool's followed its fee settings within days, but not its code upgrades.
\item A pool-level timeline of nonstandard \texttt{OP\_RETURN} use before and after \texttt{v30.0}; the one survey of the same period counts the network as a whole \cite{orangesurf2026opreturnreport}.
\end{enumerate}

The split we measure, between maintainers who write defaults and a few operators who deploy them on their own schedules, resembles what our earlier work found for Lightning node platforms \cite{kumar2026upgradedynamics}: there, too, the deployer's schedule set the pace. Section~\ref{sec:background} covers mempool policy and prior work, Sections~\ref{sec:data} and~\ref{sec:methods} the data and the estimator, Section~\ref{sec:results} the results, Section~\ref{sec:templates} template sharing, and Section~\ref{sec:discussion} what the results mean and their limits.

\section{Background and Prior Work}\label{sec:background}

\subsection{Mempool Policy, Settings and Code}
Bitcoin separates consensus, which every node enforces on blocks, from mempool policy, which each node applies to unconfirmed transactions before they reach a block. Because policy is not consensus, a pool may mine a valid transaction that most nodes would not relay \cite{b10c2024nonstandard,mollakhani2026linearization}. Most transactions, however, reach pools through the peer-to-peer network, so what nodes relay shapes what pools see.

A pool can change some rules with a startup setting. The fee floors (\texttt{-minrelaytxfee}, \texttt{-blockmintxfee}) and the \texttt{OP\_RETURN} size limit (\texttt{-datacarriersize}) are settings. Other rules are written into the code. Before \texttt{v28.0}, no setting made version-3 transactions standard, and before \texttt{v30.0} none allowed several \texttt{OP\_RETURN} outputs in one transaction; mainnet nodes refuse \texttt{-acceptnonstdtxn}. A pool that wants such a rule earlier must run newer or patched software, or take transactions outside the peer-to-peer network. MARA's Slipstream \cite{mara2024slipstream} is one such direct-submission service; private intake of this kind makes transaction ordering harder to audit \cite{messias2021selfish,messias2023dissecting}.

Pools build block templates, and the template decides which policy is in force. A few large pools appear to share templates: 0xB10C found AntPool and several other pools mining near-identical transaction sets and sending their miners the same invalid jobs \cite{b10c2024templates,b10c2025invalidjobs,b10c2026friends}. Template building can also move to individual miners. OCEAN's DATUM protocol lets its miners build their own templates \cite{hughes2024datum}; Stratum V2 job declaration lets a miner declare a template to a pool \cite{sv2jdspec}, first used for a mainnet block in June 2026 \cite{dmnd2026jd}; and P2Pool v2 has every node fetch templates from its operator's own Bitcoin node \cite{p2poolv2repo}. Where these systems run, a pool label stands for many template builders (Section~\ref{sec:limitations}).

\subsection{The 2024--2026 Wave}
Table~\ref{tab:registry} lists the rules we study and how each shipped. Four serve Lightning and other contracts that pre-sign transactions: TRUC limits a transaction's unconfirmed family so that it is harder to pin, 1p1c package relay lets a fee-paying child carry a zero-fee parent, P2A gives a keyless anchor that anyone can spend to add fees, and ephemeral dust allows one tiny output that a child spends in the same block \cite{kumar2026forceclosecosts}. Debate over lowering the default fee floor began by 2022 \cite{bitcoindev2022minrelay} and peaked in mid-2025 \cite{zhao2025minrelay,optech2025minrelaytopic}; PR~\SubOnePR{} cut it to 0.1~sat/vB \cite{core2025pr33106}. \texttt{v30.0} raised the default \texttt{OP\_RETURN} size limit and allowed several such outputs per transaction \cite{core2025v30notes,poinsot2025opreturn}. \texttt{v28.0} also turned on full replace-by-fee by default; blocks cannot show which pools accept replacements, so we measure only its node side \cite{b10c2022fullrbfnodes,todd2023fullrbf}.

\begin{table*}[t]
\centering
\caption{The Rules We Study: How Each Shipped in Bitcoin Core, and Whether a Pool Could Turn It On With a Setting}
\label{tab:registry}
\resizebox{\textwidth}{!}{% Generated by src/09_numbers.py -- do not edit by hand.
\begin{tabular}{@{}lllcccc@{}}
\toprule
Rule & Mechanism & Class & PR & Merge & First tag & Major tag \\
\midrule
Sub-1 sat/vB relay & \texttt{-minrelaytxfee=100}, \texttt{-blockmintxfee=1} & setting & \#33106 & 2025-08-15 & 2025-09-03 (v29.1) & 2025-10-09 (v30.0) \\
One OP\_RETURN $>$ 83~B & \texttt{-datacarriersize=100000} & setting & \#32406 & 2025-06-09 & 2025-10-09 (v30.0) & 2025-10-09 (v30.0) \\
Several OP\_RETURNs & Lift one-output limit in \texttt{policy.cpp} & code & \#32406 & 2025-06-09 & 2025-10-09 (v30.0) & 2025-10-09 (v30.0) \\
TRUC (version~3) & Raise \texttt{TX\_MAX\_STANDARD\_VERSION} to 3 & code & \#29496 & 2024-06-07 & 2024-10-01 (v28.0) & 2024-10-01 (v28.0) \\
1p1c package relay & Opportunistic 1p1c package validation & code$^\ast$ & \#28970 & 2024-04-30 & 2024-10-01 (v28.0) & 2024-10-01 (v28.0) \\
Ephemeral dust & One dust output if spent in package & code$^\ast$ & \#30239 & 2024-11-12 & 2025-04-11 (v29.0) & 2025-04-11 (v29.0) \\
P2A spend & Standardize \texttt{OP\_1 <0x4e73>} spends & code & \#30352 & 2024-08-02 & 2024-10-01 (v28.0) & 2024-10-01 (v28.0) \\
\midrule
OP\_RETURN 41--80~B (2015) & \texttt{-datacarriersize} default 40 $\to$ 80~B & setting & \#5286 & 2015-02-03 & 2015-07-10 (v0.11.0) & 2015-07-10 (v0.11.0) \\
Taproot spend (2021) & BIP~341/342 consensus soft fork & soft fork & \#21701 & 2021-04-16 & 2021-11-14 (act.) & 2021-04-29 (v0.21.1) \\
Full-RBF default (unobs.) & \texttt{-mempoolfullrbf=1} by default & setting & \#30493 & 2024-08-05 & 2024-10-01 (v28.0) & 2024-10-01 (v28.0) \\
\bottomrule
\end{tabular}
}
\vspace{1mm}
{\footnotesize $^{\ast}$A pool could mine the same pattern with its own fee settings, but receiving such packages from peers needed new Core code.}
\end{table*}

\subsection{Prior Measurements}
Community monitors have documented parts of this wave. 0xB10C's monitors list the first pools to mine P2A spends and ephemeral dust and track the Core versions pools run \cite{b10c2025bnocversions,b10c2026p2achart,b10c2026edchart}; an issue on 0xB10C's mainnet-observer lists pools that did not mine zero-fee packages \cite{tankyleo2025issue101}. In mid-2025 mononaut counted sub-1~sat/vB transactions and estimated the hashrate mining them \cite{scoresby2025mononaut}, and trade media reported the shift pool by pool \cite{scandizzo2025normalizan,scandizzo2025antpool,scandizzo2025foundry}. Towns \cite{towns2026opreturn}, Optech \cite{optech2026n393}, Murch \cite{murch2026dune} and orangesurf \cite{orangesurf2025defaultpolicy,orangesurf2026opreturnreport} measured \texttt{OP\_RETURN} use around \texttt{v30.0} for the network as a whole. None of them dates adoption per pool across the whole wave, separates demand from adoption, or compares miners with nodes. Section~\ref{sec:validation} checks our numbers against theirs.

\section{Data}\label{sec:data}

\subsection{Blocks and Pools}
Our main window runs from block \WindowStartHeight{} (\WindowStartDate{}) to block \WindowEndHeight{} (\WindowEndDate{}): \NBlocks{} blocks, read transaction by transaction from a local copy of the chain. For \texttt{OP\_RETURN}, where two pools moved early, we read back to block \WideStartHeight{} (\WideStartDate{}). Two control windows cover the 2015 \texttt{OP\_RETURN} change (blocks \FifteenStartHeight{} to \FifteenEndHeight{}) and Taproot's activation in 2021 (\TaprootBlocks{} blocks from \TaprootWindowStart{}).

Each block's pool comes from the public \texttt{mining-pools-data} attribution, pinned at commit \texttt{\PoolsRepoCommit{}}, which reads coinbase tags, payout addresses and co-occurring payouts. Re-resolving every block from the chain gives \PoolVerifyDiffering{} differences across \PoolVerifyBlocks{} blocks. In the main window, \PoolResolvedPctPrecise{}\% of blocks have a pool; the \PoolUnknownBlocks{} unknown blocks come from many unrelated miners, so they never count as one adopter. Sub-pools that use a host pool's templates count with the host (TrustPool with ViaBTC), and BTC.com appears under its new name, CloverPool. We draw adoption curves for the \NPoolsCurves{} pools with at least \MinBlocksPool{} blocks in the window. Unknown blocks and smaller pools stay in every denominator as not adopted, so adopted shares are lower bounds; counting them all as adopted from the start moves no $T_{50}$ date.

\subsection{Separating Lightning Anchors from Token Mints}
From \RunesBurstDate{}, token-minting chains (Alkanes and other runestone protocols) spent \PtaRunesSpends{} P2A outputs, \PtaRunesOutputPct{}\% of all P2A spends in our window. They are not the zero-fee anchors P2A was made for. We drop a P2A spend if it or the output it spends carries a runestone, which leaves \PtaLtwoSpends{} spends; \PtaLtwoTrucZero{} of them spend the anchor of a zero-fee version-3 parent, the pattern Lightning uses \cite{kumar2026forceclosecosts}. Section~\ref{sec:templates} uses the token-minting chains separately, as a tracer.

\subsection{Fees}
For every block we sum the fees paid by each rule's transactions, using the same definitions that count them. Where a zero-fee parent needs a child to pay for it (1p1c, ephemeral dust, P2A anchors), we count the child's fee, since a pool that refused the parent could not mine the child either. Our main measure is the share of fee income that a rule's transactions paid the pools that had adopted it, in the \FeeWindowDays{} days after half the hashrate adopted (Table~\ref{tab:fees}). The block subsidy (\FeeSubsidyBtc{}~BTC) is far larger than fees in this period, so we also report each rule's share of total block revenue. These are gross amounts: some users would have paid more in another form had the rule not existed, so the gain to a pool is at most this much.

\subsection{Relay Nodes}\label{sec:data-nodes}
For the node side we use the daily snapshots of KIT's Bitcoin monitor \cite{kit2026dsnmonitor}, which lists every reachable clearnet node it connects to, with the client version and service bits each announces. We read one snapshot per day from \NodeFirstDate{} to \NodeDataEndDate{}, from a frozen copy. A node relays a rule if its release does so under default settings, which we decide per release tag from the source code of Core and of Knots, whose defaults differ \cite{knots2026source}. Development builds and other clients are left out; since mid-2024 they are at most \NodeUnmappedMaxPct{}\% and \NodeOtherMaxPct{}\% of nodes. We remove large groups of nodes that announce one version from one network operator, and repeat every node date with them, with other denominators and with smoothing (\NodeVariants{} variants in all). As for hashrate, a node $T_{50}$ is the first day the share reaches one half.

\section{Method and Validation}\label{sec:methods}

\subsection{Why Counting Blocks Fails}
\begin{figure}[t]
\centering
\includegraphics[width=\columnwidth]{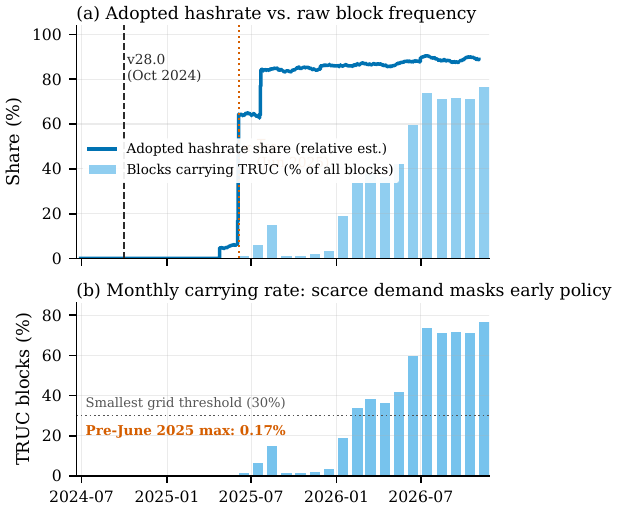}
\caption{Why demand hides adoption, for TRUC. (a) The adopted hashrate share under our estimator, against the share of all blocks carrying a TRUC transaction each month. (b) Before June 2025 the monthly share never exceeded \TrucNetShareMaxPreJunePct{}\%, far below the smallest fixed threshold in our grid.}
\label{fig:demand}
\end{figure}

A pool's blocks show a new rule only when users send transactions that need it. Core \texttt{v28.0} made TRUC transactions standard in October 2024, but until June 2025 they appeared in at most \TrucNetShareMaxPreJunePct{}\% of blocks in any month (Fig.~\ref{fig:demand}). Under a fixed threshold, say \GridSMinPct{}\% of a pool's recent blocks, every pool looks like a holdout until demand arrives. Across our sensitivity grid, fixed thresholds never date half the hashrate for ephemeral dust and P2A spends in \EphDustCellsNoTFifty{} of their \GridHalfCells{} cells (Table~\ref{tab:sensitivity}). An exploratory pass with such thresholds showed the problem; we then fixed the rule below, and its parameters, before running it on any pool.

\subsection{The Estimator}
We compare each pool with everyone else over the same blocks. A block \emph{carries} a rule if it holds at least $m$ qualifying transactions: $m = 1$ for every rule except sub-1~sat/vB, where low-fee transactions are common and $m = \FrozenMinTxsSubOne{}$. A pool adopts a rule at the first run of $W$ consecutive blocks of its own in which at least \FrozenFloor{} carry it and the share that carries it is at least half the share among all other pools' blocks over the same heights, widened by \RefPadBlocks{} blocks on each side. $W$ is \FrozenWindowRare{} blocks for the rare rules and \FrozenWindowSubOne{} for sub-1~sat/vB. The adoption date is the first carrying block of that run. For sub-1~sat/vB we count only transactions paying 0.1 to 1~sat/vB that no same-block child paid for, so a pool that merely mined a cheap parent with its child does not count.

The adopted share of hashrate on a day is the summed share of the last \HashWinBlocks{} blocks mined by pools that have adopted by then. $T_{50}$ and $T_{80}$ are the first days it reaches one half and four fifths. We measure them in days from Core's first release carrying the rule, backports included; the merge date and the next major release are reference dates (Table~\ref{tab:dates}). For a pool that never adopts, a one-sided Poisson test asks how many of its blocks would have carried the rule had it matched the other pools within \OppWinBlocks{} blocks.

We vary every parameter in a \GridCells{}-cell grid: $W$ in \{\GridWList{}\}, the required share in \{\GridSList{}\}, $m$ in \{\GridMRare{}\} (or \{\GridMHigh{}\} for sub-1~sat/vB), each with the relative comparison and with a fixed threshold. Every relative cell that reaches $T_{50}$ gives the same sign against Core's release for every rule (Table~\ref{tab:sensitivity}).

\subsection{A Test Fixed in Advance}
Before running it, we wrote down one hypothesis: that pools lead when a change needs only a setting, and wait for Core when it needs code. \texttt{v30.0}'s \texttt{OP\_RETURN} change offers a clean case, because before it one form (a single output over 83 bytes) needed only the \texttt{-datacarriersize} setting and the other (several outputs per transaction) needed new code. The hypothesis is supported if, on \texttt{v30.0}'s release day, the adopted share for the setting form is at least twice that for the code form and more pools mined it; it fails if the code form's share is at least as high.

\subsection{Validation}\label{sec:validation}
Our counts match every published record we found (Table~\ref{tab:validation}). We find the same first blocks as 0xB10C for every pool he lists as an early P2A or ephemeral-dust miner (\ValBnocPtaExact{} of \ValBnocPtaListed{} and \ValBnocEdExact{} of \ValBnocEdListed{}), the same first Taproot spend (block \ValFirstPtrHeight{}), and Towns's count of multi-\texttt{OP\_RETURN} transactions exactly. We find \ValMononautRatioPct{}\% of mononaut's sub-1~sat/vB transactions, starting from a later block, and are within a percentage point of the hashrate shares reported in the press. orangesurf's \texttt{OP\_RETURN} report names three months with spikes of nonstandard outputs before February 2026 \cite{orangesurf2026opreturnreport}; they are our three largest months, with the same causes.

Taproot is our negative control. It activated by consensus at block \TaprootActivationHeight{} (\TaprootActivationDate{}), so every pool that had upgraded could mine its spends from the same block, and an estimator that works should report no lag. Ours finds \TaprootDayZeroPools{} of the \TaprootPools{} pools adopting on day zero and $T_{50}$ on the activation day itself. The months-long lags below are therefore not an artifact of the estimator.

\begin{table*}[t]
\centering
\caption{Validation Against Published Records and the 2021 Taproot Control}
\label{tab:validation}
\resizebox{\textwidth}{!}{% Generated by src/09_numbers.py -- do not edit by hand.
\begin{tabular}{@{}llll@{}}
\toprule
Benchmark / Source & Metric & Published & This paper \\
\midrule
0xB10C BNOC \cite{b10c2025bnocversions} & First P2A / eph.-dust block heights & 8 / 7 pools & 8/8 and 7/7 exact \\
0xB10C P2TR obs.~04 \cite{b10c2022p2tr} & First block with Taproot spend & 709{,}635 & 709{,}635 (exact) \\
mononaut tx count \cite{scoresby2025mononaut} & Sub-1 sat/vB txs to block 915{,}000 & 9{,}002{,}455 & 8{,}906{,}104 (98.9\%, from 850{,}000) \\
mononaut press \cite{scandizzo2025normalizan,scandizzo2025foundry} & Adopted sub-1 share (15~Jul / 27~Jul) & $>$26\% / 85\% & 26.3\% / 85.2\% (1{,}008-blk) \\
Towns Delving \cite{towns2026opreturn} & Multi-OP\_RETURN txs (915{,}800--936{,}000) & 61 & 61 (exact) \\
F2Pool floor (vs PR~\#33106) & Strict childless min rate, 29~Jun--17~Aug & 0.50~sat/vB floor & 18{,}600 in [0.5,1), 0 in [0.2,0.5), 1 in [0.1,0.2) \\
orangesurf, mempool research \cite{orangesurf2026opreturnreport} & Nonstandard OP\_RETURN spike months before Feb.~2026 & Mar.~2024, May 2025, Aug.~2025 & same 3 months and causes \\
Taproot simultaneous control & Frozen $T_{50}$ / $T_{80}$ vs activation (2021-11-14) & Day 0 consensus & 0~d / $+$6~d (0/7{,}000 unknown) \\
\bottomrule
\end{tabular}
}
\end{table*}

\section{Results}\label{sec:results}

Who moved first differed from rule to rule. Table~\ref{tab:thesis} gives the three dates for every rule, Fig.~\ref{fig:curves} the hashrate curves, Fig.~\ref{fig:heatmap} each pool's date, and Table~\ref{tab:fees} what each rule paid.

\begin{table*}[t]
\centering
\caption{Who Moved First. For each rule: Core's first release carrying it, the day half the hashrate mined it and half of reachable nodes relayed it by default (days from Core's release; negative means before), the hashrate share at the end of the window, and the share of fee income its transactions paid adopting pools in the \FeeWindowDays{} days from the hashrate $T_{50}$.}
\label{tab:thesis}
\resizebox{\textwidth}{!}{% Generated by src/09_numbers.py -- do not edit by hand.
\begin{tabular}{@{}llllrlrrr@{}}
\toprule
& & & \multicolumn{2}{c}{Half of hashrate} & \multicolumn{2}{c}{Half of nodes} & Hashrate & Fees \\
\cmidrule(lr){4-5}\cmidrule(lr){6-7}
Rule & Change is & First Core release & date & days & date & days & at end & paid \\
\midrule
\multicolumn{9}{@{}l}{\emph{Fee floor}} \\
\quad Sub-1 sat/vB & setting & v29.1, 2025-09-03 & 2025-07-22 & $-$43 & 2026-09-06$^\dagger$ & $+$368 & 100\% & 3.1\% \\
\addlinespace
\multicolumn{9}{@{}l}{\emph{Lightning safety}} \\
\quad TRUC (version 3) & code & v28.0, 2024-10-01 & 2025-06-04 & $+$246 & 2025-07-01 & $+$273 & 89\% & 0.016\% \\
\quad 1p1c package & code$^\ast$ & v28.0, 2024-10-01 & 2025-06-14 & $+$256 & 2025-11-13 & $+$408 & 89\% & 0.016\% \\
\quad Ephemeral dust & code$^\ast$ & v29.0, 2025-04-11 & 2026-05-30 & $+$414 & 2026-08-24 & $+$500 & 89\% & 0.004\% \\
\quad P2A spend & code & v28.0, 2024-10-01 & 2026-05-30 & $+$606 & 2025-07-10 & $+$282 & 89\% & 0.004\% \\
\addlinespace
\multicolumn{9}{@{}l}{\emph{Data carrier}} \\
\quad One OP\_RETURN $>$ 83 B & setting & v30.0, 2025-10-09 & 2026-05-08 & $+$211 & 2026-11-04$^\dagger$ & $+$391 & 68\% & 0.075\% \\
\quad Several OP\_RETURNs & code & v30.0, 2025-10-09 & 2026-05-10 & $+$213 & 2026-11-04$^\dagger$ & $+$391 & 67\% & 0.024\% \\
\midrule
OP\_RETURN 41--80 B (2015) & setting & v0.11.0, 2015-07-10 & 2015-10-21 & $+$103 & --- & --- & 98\% & --- \\
\bottomrule
\end{tabular}
}
\vspace{1mm}
{\footnotesize $^{\ast}$A pool could mine the pattern with its own fee settings, but receiving it from peers needed new code. $^{\dagger}$Reached in \NodeSubOneVariantsTFifty{} of the \NodeVariants{} node variants for sub-1~sat/vB and in \NodeOpRetOneVariantsTFifty{} for \texttt{OP\_RETURN}; the node data run to \NodeDataEndDate{}, \NodeDaysPastBlocks{} days past the last block. We did not map 2015 releases to rules, so the 2015 rule has no node date.}
\end{table*}

\subsection{Fee Floor: Miners Moved First}\label{sec:res-subone}
\begin{figure*}[t]
\centering
\includegraphics[width=\textwidth]{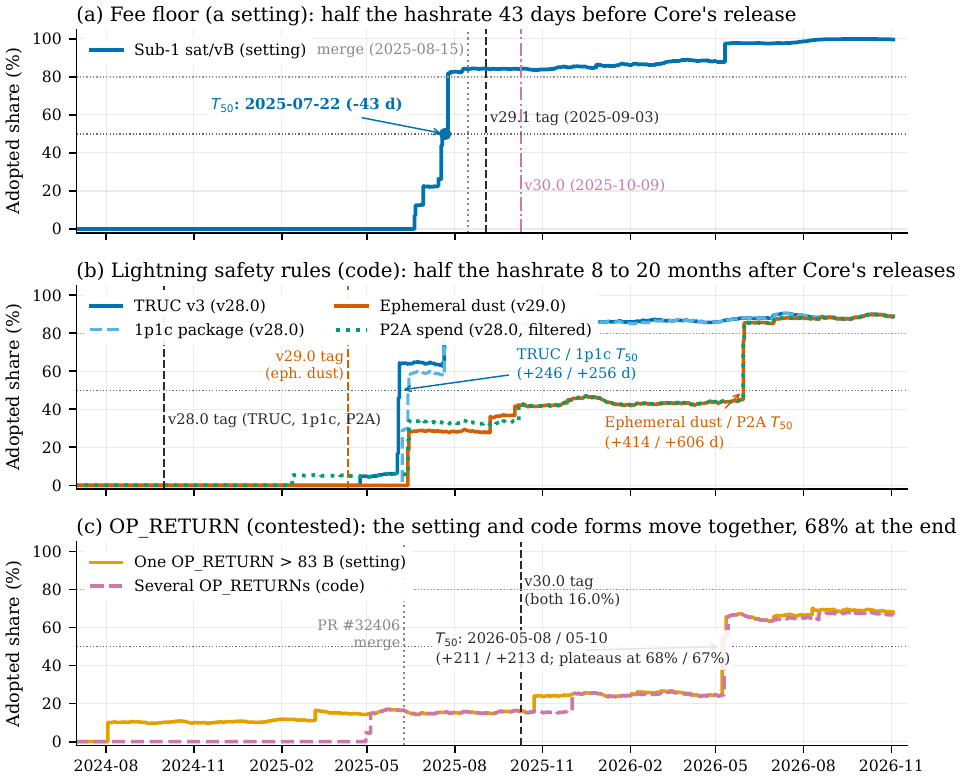}
\caption{Share of hashrate that had adopted each rule, blocks \WindowStartHeight{} to \WindowEndHeight{}. (a) The fee floor: pools moved before Core's \SubOneCoreRelease{}. (b) The Lightning safety rules: half the hashrate came \TrucDTFiftyMonths{} to \PtaDTFiftyMonths{} months after Core's releases. (c) \texttt{OP\_RETURN}: the setting form and the code form moved together.}
\label{fig:curves}
\end{figure*}

Pools cut the fee floor before Core shipped the change and before any node relayed it. Half the hashrate mined sub-1~sat/vB transactions on \SubOneTFiftyDate{}, \SubOneDTFiftyAbs{} days before \SubOneCoreRelease{} carried the lower floor (Fig.~\ref{fig:curves}a). No release relayed such transactions by default on that day; the share of nodes that did was \NodeSubOneAtHashPct{}\% on the last snapshot before it. Nodes reached half only on \NodeSubOneTFiftyDate{}, \NodeSubOneDHashAbs{} days after miners.

The change needed only two settings, and it paid. In the \FeeWindowDays{} days after half the hashrate adopted, sub-1~sat/vB transactions brought adopters \FeeSubOnePct{}\% of their fee income, or \FeeSubOneSatBlock{} satoshis per block (Table~\ref{tab:fees}). Against the block subsidy this is small, \FeeSubOneRevPct{}\% of total revenue. It is still more than \FeeSubOneTimesNext{} times the share any other rule paid.

\begin{figure*}[t]
\centering
\includegraphics[width=\textwidth]{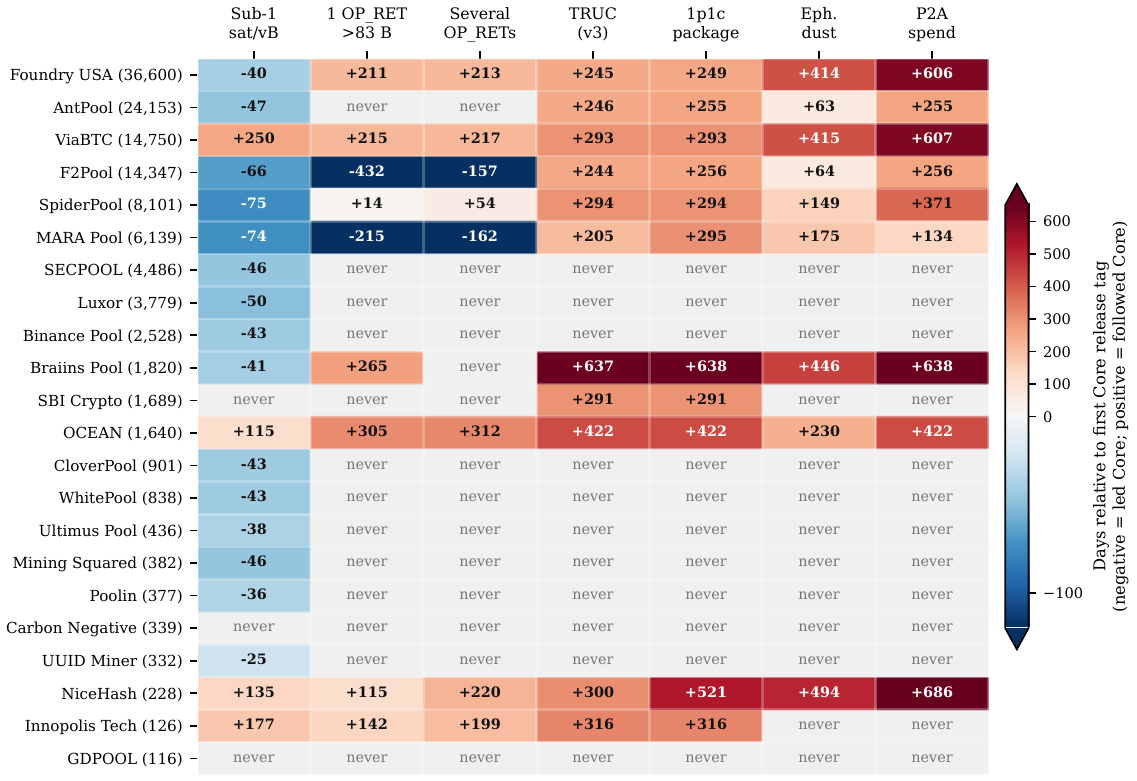}
\caption{Each pool's adoption date for each rule, in days from Core's first release carrying it, for the \NPoolsCurves{} pools with at least \MinBlocksPool{} blocks, ordered by blocks mined in the window.}
\label{fig:heatmap}
\end{figure*}

Three pools went first: SpiderPool, MARA Pool and F2Pool, all more than two months before Core's release (Fig.~\ref{fig:heatmap}). Most pools followed in mid-July 2025. The largest pools to adopt after the release were OCEAN and ViaBTC. Of the \SubOneNeverPools{} pools that never adopted, the largest was \SubOneNeverLargestPool{}, with \SubOneNeverLargestBlocks{} blocks; none of them mined a block after \SubOneNeverLastDate{}, which is why the curve ends at \SubOneAdoptedEndPct{}\%. F2Pool lowered its floor in steps. Until mid-August it mined \ValFtwoHalfTxs{} childless transactions between 0.5 and 1~sat/vB and only \ValFtwoHalfOutlierTxs{} below 0.5. It then stepped down to \ValFtwoQuarterMinRate{}~sat/vB on \ValFtwoQuarterDate{} and to 0.1~sat/vB on \ValFtwoPointOneDate{}.

\subsection{Lightning Safety Rules: Core First, Miners Months Later}\label{sec:res-code}
The four rules that protect Lightning exits reached half the hashrate \TrucDTFiftyMonths{} to \PtaDTFiftyMonths{} months after Core shipped them (Fig.~\ref{fig:curves}b). TRUC crossed on \TrucTFiftyDate{}, \TrucDTFiftyAbs{} days after \TrucCoreRelease{}, and 1p1c packages on \OnePcTFiftyDate{}. Ephemeral dust and P2A anchor spends crossed together on \PtaTFiftyDate{}, \EphDustDTFiftyAbs{} and \PtaDTFiftyAbs{} days after their releases. None of the four paid: their transactions brought adopters at most \FeeCodeMaxPct{}\% of fee income (Table~\ref{tab:fees}).

The curves move in steps because they wait on a few large pools. Foundry USA, the largest pool in the window, switched on TRUC and 1p1c in early June 2025 but ephemeral dust and P2A spends only on \PtaFoundryDate{}, and each switch moved the network curve. MARA Pool adopted TRUC on \TrucMaraDate{} and P2A spends on \PtaMaraDate{}, well before most pools. All four rules stopped at \TrucAdoptedEndPct{}\% of hashrate, because \FriendsNeverCode{} pools whose block templates resemble AntPool's never adopted them (Section~\ref{sec:templates}).

\subsection{Data Carrier: Contested, and the Setting Test Fails}\label{sec:res-d2}
Large \texttt{OP\_RETURN} outputs were contested, and our test of whether settings let pools lead fails on them. On \texttt{v30.0}'s release day, the form that needed only a setting and the form that needed new code each had \DTwoShareOnePct{}\% of hashrate (Fig.~\ref{fig:curves}c). By the decision rule we set before running the test, the hypothesis is \DTwoVerdict{}. The verdict holds in \DTwoGridNotSupported{} of the \DTwoGridCells{} grid cells, and in each cell that disagrees the difference is one pool. Both forms reached half the hashrate in May 2026, \OpRetOneDTFiftyAbs{} and \OpRetManyDTFiftyAbs{} days after release, and \OpRetOneAdoptedEndPct{}\% and \OpRetManyAdoptedEndPct{}\% by the end of the window.

Fees did not decide it either way. In the \FeeWindowDays{} days after half the hashrate adopted, large single outputs paid adopters \FeeOpRetOnePct{}\% of fee income and multi-output transactions \FeeOpRetManyPct{}\%, so pools that held back gave up almost nothing (Table~\ref{tab:fees}).

\begin{figure*}[t]
\centering
\includegraphics[width=\textwidth]{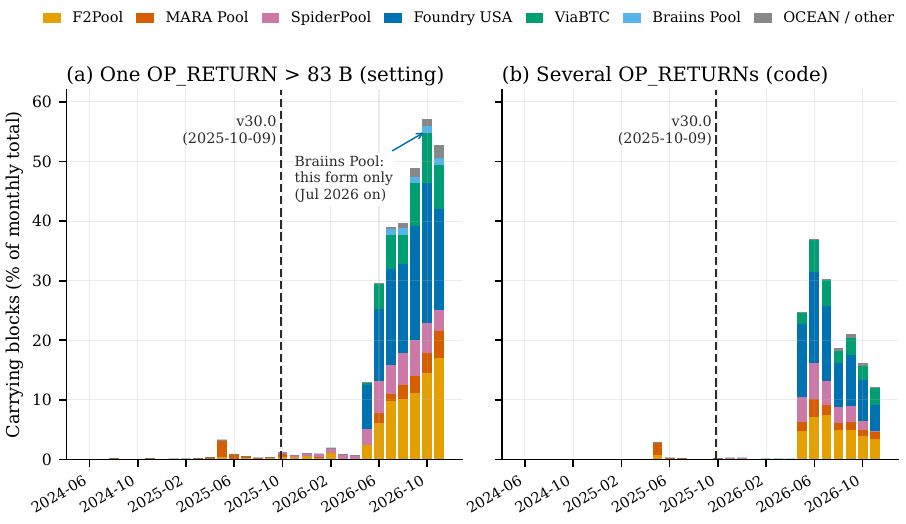}
\caption{Blocks each month carrying a large single \texttt{OP\_RETURN} output (left) or several \texttt{OP\_RETURN} outputs in one transaction (right), by pool. Before \texttt{v30.0} (\VThirtyDate{}) only F2Pool and MARA Pool mined either form in volume.}
\label{fig:opreturn}
\end{figure*}

One pattern, seen only after the test, separates the two early movers: both already mined transactions the public network did not relay. Before \texttt{v30.0}, only F2Pool and MARA Pool, which sells direct submission through Slipstream \cite{mara2024slipstream}, mined either form in volume (Fig.~\ref{fig:opreturn}). From \PreVThirtyFirstDate{} to the release they mined \PreVThirtyOneTxsDirect{} large single outputs and \PreVThirtyManyTxsDirect{} multi-output transactions. In our main window, from \WindowStartDate{} to the release, such transactions made up \FeePreVthirtyDirectPct{}\% of their fee income. F2Pool did the same in 2015 (Section~\ref{sec:res-2015}). SpiderPool followed within weeks of the release and Foundry USA and ViaBTC in May 2026, while AntPool mined neither form. Braiins Pool alone followed the setting line: from \OpRetOneBraiinsDate{} it mined large single outputs but never several in one transaction.

On-chain use followed pool adoption rather than the release: orangesurf found no rise in nonstandard \texttt{OP\_RETURN} use after \texttt{v30.0} \cite{orangesurf2026opreturnreport}, and our counts agree, with a median of \OrepAfterVthirtyMedianTxs{} such transactions a month from \OrepAfterVthirtyFirstMonth{} to \OrepAfterVthirtyLastMonth{}. After half the hashrate adopted in May 2026, the median rose to \OrepAfterTfiftyMedianTxs{} a month, with \OrepAfterTfiftyMaxTxs{} in \OrepAfterTfiftyMaxMonth{}.

Nodes held out longer than pools. When half the hashrate mined large outputs, \NodeNotRelayOpRetAtHashPct{}\% of reachable nodes did not relay them by default, and \NodeKnotsAtOpRetHashPct{}\% ran Knots, which never does. On \NodeDataEndDate{}, \NodeNotRelayOpRetEndPct{}\% still did not. Filtering by most nodes did not stop these outputs from reaching pools that would mine them.

\subsection{Nodes Moved Last on Most Rules}\label{sec:res-nodes}
\begin{figure*}[t]
\centering
\includegraphics[width=\textwidth]{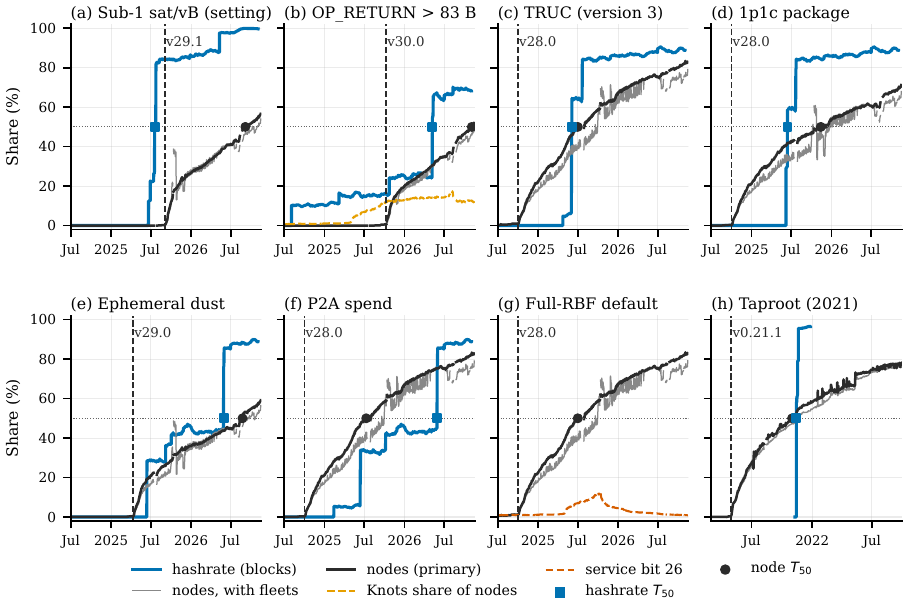}
\caption{Share of reachable clearnet nodes whose release relays each rule by default (black; gray with large host groups), beside the hashrate curves of Fig.~\ref{fig:curves} (blue). Dashed lines mark Core's first release carrying the rule; markers show each $T_{50}$. Panel (b) adds Knots' share of nodes; panel (g) the share announcing the full-RBF service bit, and has no hashrate curve because blocks cannot show the full-RBF default. In panel (h) the hashrate curve covers blocks \TaprootWindowStart{} to \TaprootWindowEnd{}. Node data: KIT \cite{kit2026dsnmonitor}, to \NodeDataEndDate{}; gaps are days without a snapshot.}
\label{fig:nodes}
\end{figure*}

On \NodeRulesMinersFirst{} of the \NodeRulesWave{} rules with their own node defaults (the two \texttt{OP\_RETURN} forms share one), half the hashrate mined the rule before half of reachable nodes relayed it by default (Fig.~\ref{fig:nodes}). On the day the hashrate crossed, between \NodeMinersFirstShareMinPct{}\% (sub-1~sat/vB) and \NodeMinersFirstShareMaxPct{}\% (TRUC) of nodes relayed the rule. Miners led by \NodeSubOneDHashAbs{} days on sub-1~sat/vB, \NodeOpRetOneDHashAbs{} on large \texttt{OP\_RETURN} outputs and \NodeOnePcDHashAbs{} on 1p1c packages, which Knots does not relay with zero-fee parents by default. The leads were shorter on ephemeral dust, \NodeEphDustDHashAbs{} days, and on TRUC, \NodeTrucDHashAbs{}. P2A spends are the exception: nodes passed half \NodePtaDHashAbs{} days before miners, because \texttt{v28.0} relays them and most nodes ran it long before Foundry mined them.

Nodes also show what blocks cannot: the full-RBF default. Nodes whose release replaces transactions without an opt-in signal passed half on \NodeFullRbfTFiftyDate{}, \NodeFullRbfDCore{} days after \texttt{v28.0}. The service bit that advertises full-RBF is a poor proxy for it: at its peak \NodeBitTwentySixMax{} nodes set it, \NodeBitTwentySixMaxKnotsPct{}\% of them Knots, and Knots stopped setting it in October 2025. At Taproot's activation, \NodeTaprootAtActivationPct{}\% of nodes already relayed its spends, and pools mined them from the first day.

\subsection{Out of Sample: the 2015 \texttt{OP\_RETURN} Change}\label{sec:res-2015}
\begin{figure}[t]
\centering
\includegraphics[width=\columnwidth]{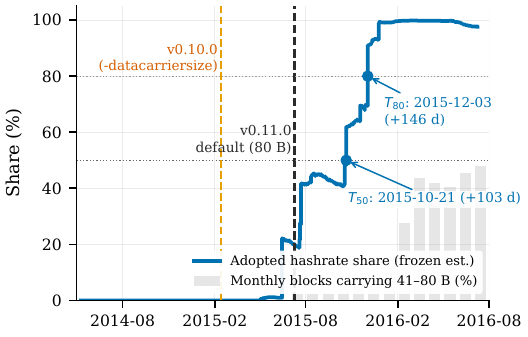}
\caption{The 2015 \texttt{OP\_RETURN} change (41--80 bytes). The setting arrived with \texttt{v0.10.0} (\FifteenSettingDate{}); adoption waited for the new default in \texttt{v0.11.0} (\FifteenDefaultDate{}).}
\label{fig:r9}
\end{figure}

The 2015 change repeats the data-carrier pattern. Core \texttt{v0.10.0} added the \texttt{-datacarriersize} setting on \FifteenSettingDate{}, and \texttt{v0.11.0} raised the default limit to 80~bytes on \FifteenDefaultDate{}. While only the setting existed, the whole network mined \FifteenOptionTxs{} transactions that needed it. Adoption came with the new default: half the hashrate on \FifteenTFiftyDate{}, \FifteenDTFiftyAbs{} days after \texttt{v0.11.0} (Fig.~\ref{fig:r9}). F2Pool again moved early: it mined \FifteenOptionFTwoTxs{} of those transactions, and the share of its blocks carrying them was \FifteenFTwoPeerRatio{} times the other pools' share.

\begin{table}[t]
\centering
\caption{What Each Rule Paid: Fees From Its Transactions as a Share of Adopting Pools' Fee Income and Total Revenue}
\label{tab:fees}
\resizebox{\columnwidth}{!}{% Generated by src/09_numbers.py -- do not edit by hand.
\begin{tabular}{@{}lrrrrrr@{}}
\toprule
& \multicolumn{4}{c}{Adopters, 90 days from $T_{50}$} & Adopters, & Network, \\
\cmidrule(lr){2-5}
Rule & blocks & of fees & of revenue & sat/block & since adoption & 90 days \\
\midrule
\multicolumn{7}{@{}l}{\emph{Fee floor}} \\
\quad Sub-1 sat/vB & 11{,}042 & 3.1\% & 0.026\% & 82{,}889 & 4.3\% & 2.6\% \\
\addlinespace
\multicolumn{7}{@{}l}{\emph{Lightning safety}} \\
\quad TRUC (version 3) & 9{,}574 & 0.016\% & $<$0.001\% & 476 & 0.039\% & 0.012\% \\
\quad 1p1c package & 9{,}655 & 0.016\% & $<$0.001\% & 469 & 0.017\% & 0.012\% \\
\quad Ephemeral dust & 11{,}067 & 0.004\% & $<$0.001\% & 85 & 0.005\% & 0.003\% \\
\quad P2A spend & 11{,}067 & 0.004\% & $<$0.001\% & 89 & 0.004\% & 0.003\% \\
\addlinespace
\multicolumn{7}{@{}l}{\emph{Data carrier}} \\
\quad One OP\_RETURN $>$ 83 B & 8{,}389 & 0.075\% & $<$0.001\% & 1{,}631 & 0.100\% & 0.049\% \\
\quad Several OP\_RETURNs & 8{,}284 & 0.024\% & $<$0.001\% & 518 & 0.232\% & 0.015\% \\
\bottomrule
\end{tabular}
}
\vspace{1mm}
{\footnotesize Gross fees; for zero-fee parents, the paying child's fee. Revenue includes the \FeeSubsidyBtc{}~BTC subsidy.}
\end{table}

\section{Who Builds the Templates?}\label{sec:templates}

The pools that 0xB10C groups with AntPool by template similarity \cite{b10c2024templates,b10c2025invalidjobs,b10c2026friends} followed AntPool on fees but not on code. All \FriendsBig{} of them with at least \MinBlocksPool{} blocks adopted the sub-1~sat/vB floor within \FriendsSubOneMaxLag{} days of AntPool. \FriendsNeverCode{} of them never adopted any of the four Lightning safety rules, although AntPool adopted TRUC on \TrucAntDate{} and the other three on \OnePcAntDate{} (Table~\ref{tab:templates}). Pool behavior thus tests the grouping, which 0xB10C drew from template contents.

The holdouts were not short of demand. Had the \FriendsNeverCodeNoLux{} holdouts other than Luxor carried TRUC transactions at the rate of the other pools, \FriendsNeverCodeTrucExpNoLux{} of their blocks would have done so; none did. With that many expected, zero cannot be chance (one-sided Poisson test). We report every total with and without Luxor, whose membership in the group is the least certain; the result does not change. The token-minting chains of Section~\ref{sec:data} give a second, independent check: they need both P2A relay and the sub-1~sat/vB floor. They appear in \RunesNonFriendsSharePct{}\% of other pools' blocks from \RunesBurstDate{} and in \RunesAntSharePct{}\% of AntPool's, but in none of the holdouts' \RunesFriendsNoLuxBlocks{} blocks.

Braiins Pool shows what an upgrade looks like from outside. Until \TrucBraiinsDate{} it mined no TRUC transaction in \BraiinsPreTrucBlocks{} blocks, where \BraiinsPreTrucExp{} were expected, while following AntPool onto the sub-1~sat/vB floor. Within two days it adopted TRUC (\TrucBraiinsLagAbs{} days after \texttt{v28.0} shipped it) alongside 1p1c packages, ephemeral dust, P2A spends and large single \texttt{OP\_RETURN} outputs. At the same time, it began mining the token chains in \RunesBraiinsSharePct{}\% of its blocks. That all five changed together fits a move to newer template software in one step.

The simplest reading is that the holdouts build templates with older software, shared or common to the group: a fee floor is a setting their operators changed, a new rule is code they did not have. For Lightning users, the rules in force in these pools' blocks are set by when that software is updated, not by Core's release dates.

\begin{table*}[t]
\centering
\caption{The Pools Grouped With AntPool by Template Similarity: Fee Floor Followed Within Days, Code Rules Not Adopted (Shown With and Without Luxor)}
\label{tab:templates}
\resizebox{\textwidth}{!}{% Generated by src/09_numbers.py -- do not edit by hand.
\begin{tabular}{@{}lrrrrrrr@{}}
\toprule
Pool & Blocks & Sub-1 lag & TRUC & 1p1c / dust / P2A & TRUC blocks before & \multicolumn{2}{c}{Runes P2A (Jul+ 2026)} \\
 & (850k--974k) & (days) & from & from & adoption, obs / exp & carrying / total & share \\
\midrule
AntPool (reference) & 24{,}153 & 0 (2025-07-18) & 2025-06-04 & 2025-06-13 & 0 / 10.5 & 1{,}472 / 3{,}511 & 41.9\% \\
\midrule
Binance Pool & 2{,}528 & $+$4 & never & never & 0 / 440.0 & 0 / 279 & 0.0\% \\
Braiins Pool & 1{,}820 & $+$6 & 2026-06-30 & 2026-07-01 & 0 / 168.9 & 227 / 285 & 79.6\% \\
CloverPool & 901 & $+$4 & never & never & 0 / 192.2 & 0 / 207 & 0.0\% \\
Luxor$^{\dagger}$ & 3{,}779 & $-$3 & never & never & 0 / 847.8 & 0 / 644 & 0.0\% \\
Mining Squared & 382 & $+$1 & never & never & 0 / 23.9 & 0 / 0 & --- \\
Poolin & 377 & $+$11 & never & never & 0 / 57.8 & 0 / 46 & 0.0\% \\
SECPOOL & 4{,}486 & $+$1 & never & never & 0 / 1{,}063.6 & 0 / 749 & 0.0\% \\
Ultimus Pool & 436 & $+$9 & never & never & 0 / 57.6 & 0 / 17 & 0.0\% \\
\midrule
Holdouts (7, with Luxor) & 12{,}889 & $-$3..$+$11 & 0 / 7 & 0 / 7 & 0 / 2{,}682.9 & 0 / 1{,}942 & 0.0\% \\
Holdouts (6, excl.\ Luxor) & 9{,}110 & $+$1..$+$11 & 0 / 6 & 0 / 6 & 0 / 1{,}835.2 & 0 / 1{,}298 & 0.0\% \\
\bottomrule
\end{tabular}
}
\vspace{1mm}
{\footnotesize $^{\dagger}$Luxor's membership in the group is the least certain; the summary rows therefore show both cohorts. CloverPool includes blocks tagged BTC.com.}
\end{table*}

\section{Discussion}\label{sec:discussion}

\subsection{Who Sets Bitcoin's Mempool Policy?}
Core writes the defaults, but pools decide when a rule takes effect. Where a change needed only a setting and brought fees, pools set the policy before Core shipped it and before nodes relayed it. Where a change needed new code and brought almost no fees, Core's release started a clock that ran \TrucDTFiftyMonths{} to \PtaDTFiftyMonths{} months, and the template software of a few large pools set its pace. Where a change was contested, neither a setting nor money decided. Two pools that already mined transactions the public network did not relay moved early, most waited for Core, and a large group never moved.

The fee measurements sharpen what ``pays'' means. The fee floor paid modestly, \FeeSubOnePct{}\% of adopters' fee income and \FeeSubOneRevPct{}\% of their revenue, but it cost nothing to switch on. The safety rules earned pools almost no fees, so no pool had a fee reason to upgrade early; they reached each pool with its next upgrade. For \texttt{OP\_RETURN}, neither side faced a fee penalty: pools that held back lost nothing and pools that adopted gained little, which is why we read the choice as a stance rather than a revenue decision.

The test we set before the analysis failed: a setting alone did not let most pools move first. The account we give of the contested case, that pools with their own transaction intake move early, came after we saw the data. It is a hypothesis that fits two pools in 2024--2025 and one in 2015, not a finding.

Nodes did not hold miners back. On \NodeRulesMinersFirst{} of the \NodeRulesWave{} rules with their own node defaults, pools mined a rule before most nodes relayed it, so relay by a minority of nodes, or direct submission, appears to have been enough to reach the pools that wanted the transactions. The reverse held for large \texttt{OP\_RETURN} outputs: most nodes filtered them, and pools with most of the hashrate mined them anyway.

\subsection{What It Means}
\paragraph*{For Lightning and wallet developers} A Core release date is not a deployment date. Zero-fee exits that rely on P2A anchors could reach less than half the hashrate until May 2026, \PtaDTFiftyMonths{} months after \texttt{v28.0}, and \PtaAdoptedEndPct{}\% at the end of our window \cite{kumar2026forceclosecosts}. The number to watch is the share of hashrate that mines a pattern, and it moves when a few large pools upgrade.

\paragraph*{For Core maintainers} A change that pools can make with a setting reached most of them in weeks, before the release that carried it. A change that needs new code reaches pools only as fast as they adopt new releases, and several pools had not adopted \texttt{v28.0}'s rules two years after it shipped.

\paragraph*{For the \texttt{OP\_RETURN} debate} Node defaults did not predict what was mined in either direction. Pools mined sub-1~sat/vB transactions that no default relayed, and mined large \texttt{OP\_RETURN} outputs while most nodes filtered them. What reached blocks depended on which pools would mine a transaction and whether it could reach them.

\paragraph*{Beyond pools} Systems that let miners build their own templates---DATUM, Stratum V2 job declaration and P2Pool v2 \cite{hughes2024datum,sv2jdspec,p2poolv2repo}---would move this choice from pool operators to individual miners. In our window they were too small to measure as separate adopters.

\subsection{Limitations}\label{sec:limitations}
\emph{Blocks show adoption, not refusal.} Our dates rest on comparing each pool with the others over the same blocks; a rule that almost no one used cannot be dated, and we do not date full replace-by-fee from blocks at all. Transactions a pool took privately look the same on chain as relayed ones, so early adoption by pools with their own intake mixes the two.

\emph{A pool label is not always the template builder.} OCEAN's miners can build their own templates with DATUM \cite{hughes2024datum}, and OCEAN offered several template policies until 21 December 2025 \cite{ocean2025templates}, so its dates combine many builders' choices. Two blocks from the original P2Pool fall in our window and count among the unknown blocks. The AntPool group comes from public template comparisons \cite{b10c2024templates,b10c2026friends}, not from the pools' own infrastructure.

\emph{Fees are gross.} Some users would have paid more in another form had a rule not existed, so a rule's fees bound a pool's gain from above. The \FeeWindowDays{}-day window is our choice; Table~\ref{tab:fees} also gives each rule's share since each pool's adoption and across all pools, and sub-1~sat/vB leads by more than ten times in every view.

\emph{Nodes.} The node series counts the reachable clearnet nodes of one monitor (\NodeEndWithUa{} on the last day); nodes that do not accept connections, Tor-only nodes and pools' own nodes are invisible, and client versions are self-reported. A default is not a setting: an operator who lowered \texttt{-minrelaytxfee} by hand relayed sub-1~sat/vB transactions before any release did. On \NodeXcheckDates{} dates KIT reads new releases \NodeXcheckKitHighPp{} points higher than btcnodes.io \cite{btcnodes2026} and bitnod.es \cite{bitmex2026bitnodes}, mostly because of one group of nodes the others count and KIT does not; without it the gap is \NodeXcheckExKitHighPp{} points. Node dates may therefore run slightly early, which would shorten the gaps by which miners led.

\emph{Testing.} We fixed the estimator and the setting test in an internal method note before running them, not in a public registry. The Poisson tests are not corrected for multiple comparisons; they rest on zero observed against hundreds or thousands expected.

\section{Conclusion}\label{sec:conclusion}

We dated each of the 2024--2026 changes to Bitcoin's mempool policy at three points: Core's release, half the hashrate and half of reachable nodes. Who set the policy depended on whether a change needed only a setting and whether it paid. Pools cut the fee floor \SubOneDTFiftyAbs{} days before Core shipped it, because a setting allowed it and it paid. They took \TrucDTFiftyMonths{} to \PtaDTFiftyMonths{} months to mine the rules that protect Lightning exits, which needed new code and earned almost no fees, and the holdouts whose templates resemble AntPool's never did. On large \texttt{OP\_RETURN} outputs, pools split: neither the setting nor fees decided. Nodes came last on most rules. A rule written into Core is in force for users only once the pools that build most templates run software that has it.

\section*{Data and Code Availability}
The code, every aggregate behind the numbers in this paper and the per-pool adoption dates (\nolinkurl{data/pool_adoption_dates.csv}) are at \url{https://github.com/bitcoindatalabs/mempool-policy-adoption}. Pool attribution is \texttt{mining-pools-data} at commit \texttt{\PoolsRepoCommit{}}; node data are KIT's Bitcoin monitor snapshots (CC BY 4.0) \cite{kit2026dsnmonitor}.

\bibliographystyle{IEEEtran}
\bibliography{references}

\appendices
\section{Sensitivity, Reference Dates and Sources}\label{sec:appendix}

\subsection{Sensitivity}
Table~\ref{tab:sensitivity} runs every rule through the \GridCells{}-cell grid of Section~\ref{sec:methods}: \GridHalfCells{} cells with the relative comparison and \GridHalfCells{} with a fixed threshold. Every relative cell that reaches $T_{50}$ gives the same sign against Core's release, for every 2024--2026 rule and for the 2015 change. Fixed thresholds often never reach $T_{50}$ when demand is scarce, which is why we use the relative comparison. Counting every unknown block and every small pool as adopted from the start raises the adopted share by at most \NonUnitMaxSharePct{}\% and moves no $T_{50}$.

\begin{table*}[t]
\centering
\caption{Sensitivity: Each Rule Across the \GridCells{} Grid Cells (\GridHalfCells{} Relative, \GridHalfCells{} Fixed-Threshold)}
\label{tab:sensitivity}
\resizebox{\textwidth}{!}{% Generated by src/09_numbers.py -- do not edit by hand.
\begin{tabular}{@{}lcccccc@{}}
\toprule
Rule & Frozen $(W, s, m)$ & Rel.\ $T_{50}$ & Same sign & Rel.\ $T_{50}$ range & Abs.\ $T_{50}$ & Status \\
\midrule
Sub-1 sat/vB & $(20, 0.5, 100)$ & 27/27 & 27/27 & 2025-07-19..2025-07-24 & 27/27 & robust \\
One OP\_RETURN $>$ 83 B & $(50, 0.5, 1)$ & 27/27 & 27/27 & 2026-05-08..2026-05-11 & 27/27 & robust \\
Several OP\_RETURNs & $(50, 0.5, 1)$ & 27/27 & 27/27 & 2026-05-10..2026-05-15 & 26/27 & robust \\
TRUC (version 3) & $(50, 0.5, 1)$ & 27/27 & 27/27 & 2025-06-04..2025-08-02 & 27/27 & robust \\
1p1c package & $(50, 0.5, 1)$ & 27/27 & 27/27 & 2025-06-14..2025-08-04 & 24/27 & robust \\
Ephemeral dust & $(50, 0.5, 1)$ & 27/27 & 27/27 & 2026-05-30..2026-09-17 & 16/27 & robust \\
P2A spend & $(50, 0.5, 1)$ & 27/27 & 27/27 & 2026-05-30..2026-09-27 & 16/27 & robust \\
P2A spend, unfiltered & $(50, 0.5, 1)$ & 27/27 & 27/27 & 2026-05-29..2026-05-30 & 27/27 & robust \\
OP\_RETURN, all & $(50, 0.5, 1)$ & 27/27 & 27/27 & 2026-05-08..2026-05-11 & 27/27 & robust \\
\midrule
OP\_RETURN 41--80~B (2015) & $(50, 0.5, 1)$ & 27/27 & 27/27 & 2015-10-21..2016-02-07 & 20/27 & robust \\
\bottomrule
\end{tabular}
}
\end{table*}

\subsection{Reference Dates}
Table~\ref{tab:dates} gives $T_{50}$ against all three Core dates, and $T_{80}$. Measured from the merge, every lag is longer; measured from the next major release, the fee floor's lead grows to \SubOneDTFiftyMajorAbs{} days.

\begin{table}[t]
\centering
\caption{$T_{50}$ Against Core's First Release, Merge and Next Major Release, and $T_{80}$}
\label{tab:dates}
\resizebox{\columnwidth}{!}{% Generated by src/09_numbers.py -- do not edit by hand.
\begin{tabular}{@{}lrrrr@{}}
\toprule
& \multicolumn{3}{c}{$\Delta T_{50}$ (days) vs} & $\Delta T_{80}$ \\
\cmidrule(lr){2-4}
Rule & first tag & merge & major & (days) \\
\midrule
Sub-1 sat/vB & $-$43 & $-$24 & $-$79 & $-$40 \\
One OP\_RETURN $>$ 83 B & $+$211 & $+$333 & $+$211 & never \\
Several OP\_RETURNs & $+$213 & $+$335 & $+$213 & never \\
TRUC (version 3) & $+$246 & $+$362 & $+$246 & $+$294 \\
1p1c package & $+$256 & $+$410 & $+$256 & $+$295 \\
Ephemeral dust & $+$414 & $+$564 & $+$414 & $+$415 \\
P2A spend & $+$606 & $+$666 & $+$606 & $+$607 \\
\midrule
OP\_RETURN 41--80 B (2015) & $+$103 & $+$260 & $+$103 & $+$146 \\
\bottomrule
\end{tabular}
}
\end{table}

\subsection{Source Checks}
Each classification in Table~\ref{tab:registry} comes from the source code at the release tags of \texttt{bitcoin/bitcoin}:
\begin{itemize}\raggedright
\item \textbf{Code.} TRUC: \texttt{TX\_MAX\_\-STANDARD\_\-VERSION} rose from 2 to 3 in \texttt{src/policy/policy.h} (\texttt{v27.0} against \texttt{v28.0}). P2A spends: \texttt{v28.0} stopped rejecting spends of the pay-to-anchor witness program as an upgradable witness version (\texttt{src/policy/policy.cpp}). Several \texttt{OP\_RETURN} outputs: the \texttt{multi-op-return} check in \texttt{src/policy/policy.cpp} was removed in \texttt{v30.0}, and mainnet refuses \texttt{-acceptnonstdtxn} (\texttt{src/node/mempool\_args.cpp}).
\item \textbf{Settings.} The fee floor: \texttt{DEFAULT\_\-MIN\_\-RELAY\_\-TX\_\-FEE} fell from 1000 to 100 satoshis per kilo-virtual-byte, and \texttt{-minrelaytxfee}, \texttt{-incrementalrelayfee} and \texttt{-blockmintxfee} accept lower values. A single large \texttt{OP\_RETURN} output, and the 2015 change, are governed by \texttt{-datacarriersize} (\texttt{v0.10.0} \texttt{src/init.cpp}).
\item \textbf{Code, but a pool could mine the pattern.} 1p1c packages (\texttt{Find1P1CPackage} in \texttt{src/net\_\-processing.cpp}) and ephemeral dust (\texttt{PreCheckEphemeralTx} in \texttt{src/policy/ephemeral\_\-policy.h}) shipped as new relay code in \texttt{v28.0} and \texttt{v29.0}; a single pool could still mine the same pattern earlier with its own fee settings.
\end{itemize}

\end{document}